\documentclass[10pt,twocolumn]{article}
\usepackage{amsmath, latexsym,
amssymb, amscd, amsfonts, amsthm}
\usepackage[english]{babel}

\DeclareFixedFont{\sfracFont}{U}{euf}{b}{n}{7pt}
\newtheorem{lem}{Lemma}[]

\newtheorem{thm}{Theorem}[]

\newcommand{\BB}[1]{\mathbb{#1}}

\newcommand{\AC}{\mathcal{A}}
\newcommand{\BC}{\mathcal{B}}
\newcommand{\CC}{\mathcal{C}}

\newcommand{\FC}{\mathcal{F}}
\newcommand{\GC}{\mathcal{G}}
\newcommand{\HC}{\mathcal{H}}

\newcommand{\LC}{\mathcal{L}}

\newcommand{\WC}{\mathcal{W}}

\begin{document}

\title{Optimal Quantum Filtering and Quantum Feedback Control}
\author{Simon C. Edwards, Viacheslav P. Belavkin \\
School of Mathematical Sciences\\
University of Nottingham\\
Nottingham. NG7 2RD. UK\\
pmxsce@nottingham.ac.uk\\
vpb@maths.nottingham.ac.uk}
\date{}
\maketitle

\begin{abstract}
Quantum mechanical systems exhibit an inherently
probabilistic nature upon measurement.  Using
a quantum noise model to describe the stochastic
evolution of the open quantum system and working
in parallel with classical indeterministic
control theory, we present the theory of
nonlinear optimal quantum feedback control.  The
resulting quantum Bellman equation is
then applied to the explicitly solvable
quantum linear-quadratic-Gaussian (LQG)
problem which emphasizes
many similarities with the corresponding
classical control problem.
\end{abstract}

\section{Introduction}

With technological advances now allowing the possibility of
continuous monitoring and rapid manipulations of systems at
the quantum level \cite{AASDM02,GSDM03},
there is an increasing awareness of the
applications and importance of quantum feedback control.
Such applications include the engineering of quantum states,
stability theory, quantum error correction and substantial
applications in quantum computation \cite{ADL02,AWM03,GrW04,SAJM04,HJHS03,SROKW02}.
This current
interest marks quantum control theory as a highly rewarding
branch of control theory for study and as such there is
a growing number of recent publications on the subject
\cite{DoJ99,DJJ01,EdB03,BEB05,Jam04,HSM05}.  In
particular, \cite{HSM05} contains a useful introduction
to quantum probability and along with \cite{DHJMT00} gives
a comprehensive discussion on the comparison of
classical and quantum control techniques and we
refer the unfamiliar reader to these articles and references within.

The main ingredients of quantum control are essentially the same
as in the classical case.  One controls the system by
coupling to an external control field which modifies the
system in a desirable manner.  The desired objectives of
the control can be encoded into a \emph{cost function}
along with any other stipulations or restrictions on the
controls such that the minimization of this cost
indicates optimality of the control process.
There are two types of dynamical control - \emph{open loop} (or
blind) control where the controls are predetermined at the start
of the experiment and \emph{closed loop} (or feedback) control
where controls can be chosen throughout the experiment and thus is
preferable for stochastic dynamics. Previous work on the theory of
optimal quantum open loop control includes variational techniques on
closed qubit systems \cite{PK02,TV02}, which was also
extended to open (dissipative) quantum systems \cite{XYOFR04}.
However, this approach can only seek locally optimal solutions
which can often be improved further with measurement and feedback,
since an open quantum system inevitably loses
information to its surrounding environment.

Quantum feedback control was formally initiated by Belavkin in a series of
papers \cite{Bel79,Bel83,Bel88} in the 80s.  This work was developed as a
quantum analogy to the classical theories of nonlinear
(Stratonovich) filtering and Bellman dynamic programming.  In
fact, the separation lemma of classical control theory was
shown also to hold in the quantum domain.  That is, the
problem of optimal quantum feedback control is separated
into quantum filtering which provides optimal estimates
of the stochastic quantum variables (operators) and
then an optimal control problem based on the output of the
quantum filter.  The quantum noise which we filter out
comes from the disturbance to the system due to
the quantum measurement.  Unlike classical systems, this is
an unavoidable feature of quantum measurement since
the quantum system is not directly observable.  The quantum filter
describes a classical stochastic process, albeit
on the space of quantum states, so Belavkin showed how one
can progress using tools from classical feedback control theory
when applied to sufficient coordinates of the system \cite{Bel83}.
However, the lack of urgency for such a
theory and the complexity of the mathematical language
at the time left this work relatively undiscovered
only to be rediscovered recently in the physics and
engineering community.

The purpose of this paper is to build on the original work
of Belavkin and present an accessible account of the theory
of nonlinear optimal quantum feedback control.  Firstly we
introduce the necessary concepts from modern quantum theory
including quantum probability, non-demolition measurement,
quantum stochastic calculus and quantum filtering.
Next the quantum Bellman equation for optimal
feedback control with diffusive non demolition measurement is derived.
Often in optimal control
problems of this nature, the separation lemma is assumed
and the control objectives are defined in terms of posterior
sufficient coordinates \cite{DHJMT00,EdB03,BEB05}.  In this paper, we show how the
general Bellman equation is applied with the same effect
by application to the many dimensional
quantum LQG problem.  Next a physcial example of LQG control
is given and we conclude with a discussion on
the results with comparison to the corresponding classical
control problem.

\section{Optimal quantum measurement and filtering}

This section highlights the differences between quantum and classical
systems and introduces the problem of quantum measurement.  After the
appropriate setting is given, the measurement problem is then restated as
a problem of optimal estimation of the output of a noisy quantum channel.
Finally, the quantum filtering equation describing the dynamical least
squares estimator is given.

\subsection{Quantum Probability}

Quantum physics which deals with the unavoidable random
nature of the microworld requires a new, more general,
noncommutative probability theory than the classical one
based on Kolmogorov's axioms.  It was developed through the
70s and 80s by Accardi, Belavkin, Gardiner, Holevo, Hudson and
Parthasarthy \cite{AFL90,Bel79,GaZ00,Hol82,Par92} amongst others.

The essential difference between classical and quantum probability
is that classically, Kolmogorov's probability axioms allow
the occurrence of simultaneous events only.  This
is because the classical events
are described by indicator functions $1_\Delta(\omega)$
of the measurable subsets $\Delta\subseteq\Omega$
on the space of point states $\Omega$. They are
the building blocks for the classical random variables described by
measurable functions $x:\Omega\to\mathbb{R}$
as linear combinations (integrals) of the indicator functions
$1_\Delta$.  Such classical essentially bounded variables
represented by operators of multiplication by the corresponding
functions, form an abelian (commutative) von Neumann algebra
on the Hilbert space $L^2(\Omega,\mathbb{P})$ of square-integrable
random functions with respect to a probability measure $\mathbb{P}$.

In quantum probability, there are some events which cannot
occur simultaneously, so we must generalize the framework
of classical probability to incorporate these features.  This
is done by considering quantum events as
self adjoint orthoprojectors $P^2=P=P^*$
(where $*$ denotes the Hermitian adjoint) acting in some Hilbert space $\HC$
not only by multiplications on the indicator functions $1_\Delta(\omega)$.
Quantum random variables are also built from events
as linear (integral) combinations of their projectors $P$.
The events are incompatible if the corresponding projectors do not
commute, i.e.\ $[P_i,P_j]:=P_iP_j-P_jP_i\ne0$ and therefore
cannot be represented classically by the indicator
functions which always commute.

One can form the non commutative von Neumann algebra $\AC$ of
bounded quantum random
variables generated by the
self adjoint projectors $\{P^1,..,P^m\}$. This algebra
is equal to its double commutant
$\AC:=\{P^1,...,P^m\}''$ where the commutant
of a set $S\subset\BC(\HC)$ in the algebra
$\BC(\HC)$ of all bounded operators on $\HC$ is defined by
$S':=\{X'\in\BC(\HC)\textrm{ s.t. }[X,X']=0$ $\forall X\in S\}$.
The quantum state on $\AC$, given by a
positive operator $\rho=\rho^*\ge0$ with unit
trace $\textrm{Tr}[\rho]=1$, defines all
expectations
\begin{equation}
\langle X\rangle=\textrm{Tr}[\rho X]=\langle\rho,X \rangle
\end{equation}
for operators $X\in\AC$.  So we describe a
quantum probability space by the pair $(\AC,\rho)$.  In the case
where $\AC$ is an abelian von Neumann operator algebra, there
is a natural isomorphism $(\AC,\rho)\simeq L^\infty(\Omega,\mathbb{P})$
with bounded functions on the classical probability space $(\Omega,\mathbb{P})$
and so we recover the classical statistics.

The incompatibility of quantum events means that
after one has observed an event, the state of the
system needs to be updated to account for the
change to the system or \emph{back-action}
affecting the expectations of all other incompatible events.
This state change was traditionally described by the
normalized projection postulate
\begin{equation}
\rho\to\rho_i=\frac{P_i\rho P_i}{\textrm{Tr}[\rho P_i]}
\end{equation}
which also ensures instantaneous repeatability
of the observed event corresponding to the projection $P_i$.
However, it has long been known that this phenomological
description is inadequate, since it fails to describe
continuous measurements and experimentally it is not possible
to perform a direct measurement of eigenstates of such a quantum
operator.  Instead we must
consider an indirect measurement of operators in a coupled
semi-classical field and describe the state change $\rho\to\rho_i$
by an optimal estimator based on the results of measurements in this field.
Let $\FC$ denote the Hilbert space of the field, which we view
as a noisy measurement channel in the initial vacuum state $\phi$.  We
only observe compatible events in the channel (corresponding to output
meter readings for example).  So we describe these
events by commuting
projectors $\{P_\omega\}_{\omega\in\Omega}$ which can be represented
by classical indicator functions and generate the abelian subalgebra
$\BC\subset\BC(\FC)$ where $\Omega$ is now the space of measurement results
(eigenvalues) for these commuting operators.
So the field operators $W\in\BC$ which are linear combinations
of the commuting projectors are in one-to-one correspondence
with classical random variables as functions
$w:\Omega\to\mathbb{R}$ on the data space $\Omega$.  In the quantum
noise model, we consider input quantum noises as quantum random
variables represented by operators in the full field algebra $\BC(\FC)$
of bounded operators on $\FC$ which perturb the quantum system in
such a way to allow a classical correlated output.
This interaction between the open quantum system
and the semi-classical field is described on the
composite system by
a unitary operator $U$, which for an initial state $\phi$ of the field
gives the state evolution
\begin{displaymath}
\rho\to U(\rho\otimes\phi)U^*
\end{displaymath}
called the \emph{prior state}.
The reduced conditional evolution can then be described by the nonlinear
map
\begin{equation}\label{eq poststate}
\rho\to\rho_\omega\:=\frac{\textrm{Tr}_\FC[U(\rho\otimes\phi)U^*(I\otimes
P_\omega)]}{\textrm{Tr}[U(\rho\otimes\phi)U^*(I\otimes P_\omega)]}
\end{equation}
called the \emph{posterior state} which is the Bayes
law of conditioning for the measurement result $\omega\in\Omega$,
normalized with respect to the output probabilities
$\mathbb{P}(\omega)=\textrm{Tr}[U(\rho\otimes\phi)U^*(I\otimes P_\omega)]$
and $\textrm{Tr}_\FC$ denotes the partial trace over $\FC$.

We denote the posterior state as a classical random variable
$\rho_\bullet:\Omega\to\AC_*$ taking values $\rho_\omega$
in the space $\AC_*$ of states on $\AC$.
The posterior state gives the conditional expectation
\begin{displaymath}
\mathbb{E}[X'|Y]=\langle\rho_\bullet,X\rangle
\end{displaymath}
which is the least squares estimator of the system operator $X'=U^*(X\otimes I)U$
after interaction,
with respect to the output operators $Y:=U^*(I\otimes W)U$.
We now describe
the appropriate model for the dynamical
coupling between the open quantum system and the
field.

\subsection{The Quantum Vacuum Noise Model and Markov Approximation}

The indirect measurement of the quantum system is
via a coupled measurement channel, playing the
role of a quantum noise bath.
It is modelled by the symmetric Fock space
$\FC$ over the single particle space
$L^2(\BB{R}_+\to \GC)$ of square integrable
functions from $[0,\infty)$
into a Hilbert space $\GC$ of
the bath degrees of freedom.
Having in mind the vacuum noise model of the bath,
let $\WC:=\BC(\FC)$ denote the quantum noise algebra
of bounded operators on $\FC$ initially in the vacuum
state $\phi$. From the
divisibility property of the symmetric Fock space, we can
factorize the noise
algebra
\begin{displaymath}
\WC=\WC_0^t\otimes\WC_t^\infty,
\qquad \FC=\FC_{[0,t)}\otimes\FC_{[t,\infty)}
\end{displaymath}
for arbitrary $t>0$ where
$\WC_a^b=\BC(\FC_{[a,b)})$ and
$\FC_{[a,b)}$ is the symmetric Fock
space over $L^2([a,b)\to \GC)$ for
$0\le a<b$. This tensor independence
implies compatibility for operators
belonging to the disjoint time
intervals of the noise algebra.
The time evolution of the quantum system
and the quantum noise bath (which together
form a closed composite quantum system)
can be described in the interaction representation
by a family
$\{U_t\}_{t\in\mathbb{R}_+}$ of unitary operators
$U_t:\HC\otimes\FC_{[0,t)}\to\HC\otimes\FC_{[0,t)}$.
In the weak coupling limit
\cite{AFL90},\cite{AGL95} (short bath memory), they
describe the Markovian \emph{flow}
$j_t:\AC\to\AC\otimes\WC_0^t$
by $j_t(X):=U_t^*(X\otimes I)U_t$
for operators $X\vdash \AC$ (we use the symbol
$X\vdash \AC$ to denote that $X$ is an element of $\AC$, or
that its spectral projectors belong in
$\AC$ for the case of unbounded $X$).
We complete the description of the joint system and field
evolution by introducing the unitary shift operator
$S_t:\FC_{[0,s)}\to\FC_{[t,s+t)}$ which models
the free evolution in the field.
Thus the combined evolution and interaction on the composite system
is given by a family of endomorphisms $\{\gamma_t\}$ on $\AC\otimes\WC$
such that
$\gamma_t(X\otimes W)=\hat{U}^*_t(X\otimes W)\hat{U}_t$ for
unitaries $\hat{U}_t:=(I\otimes S_t)U_t$. This gives the
cocyle identity $U_{t+s}=S_{-s}U_tS_sU_s$
for the interaction unitaries $\{U_t\}$.  Note that
for ease of presentation, we avoid the repetition
of tensoring with the identity on $\HC$ and $\FC_{[t,\infty)}$
and assume the domain
of the operators is clear from the context.

We now briefly
discuss quantum stochastic calculus, a necessary tool
when developing a time-continuous theory of
quantum stochastic evolution.

\subsection{Quantum Stochastic Calculus}

In this paper we consider feedback control based on a
homodyne detection scheme.  This is the quantum
analogue of measurement of the Wiener process in the field
and is described by the field quadrature $W_t=A_t+A_t^*$
where $A_t\vdash\WC_0^t$ is called the \emph{annihilation} operator
on $\FC$. The properties of $A_t$ are such that
$W_t^\theta:=\exp (\textrm{i}\theta) A_t+\exp (-\textrm{i}\theta)A^*_t$ is
equivalent to the classical Wiener process for each
$\theta\in [0,2\pi)$, however they do not commute for
different $\theta$, so by considering solely the
measurement of $W_t$, we restrict ourselves
to a chosen classical diffusive measurement process
corresponding to $\theta=0$.

Hudson and Parthasarathy \cite{HuP84},\cite{Par92}
developed the theory
of quantum stochastic calculus using the annihilation
process and its adjoint, the creation process
$A^*_t$ as the fundamental diffusive adapted
processes and defined the interaction unitaries $\{U_t\}$
as the unique solutions to the quantum stochastic
differential equation which we chose of the simple form
\begin{equation}\label{eq unitary}
dU_t+KU_t\otimes dt=LU_t\otimes dA^*_t-L^*U_t\otimes dA_t.
\end{equation}
with $U_0=I$. Here $K=\frac{\textrm{i}}{\hbar}H+\frac{1}{2}L^*L$, $H$ is the Hamiltonian
of the quantum system and $L$ is the operator
describing the coupling of the system to the measurement
channel.  The increments $dt$, $dA_t$, $dA^*_t$
are considered as operators acting in $\FC_{[t,t+dt)}$ and define
stochastic It\^o calculus using the
product rule
\begin{equation*}
d(M_tN_t)=d(M_t)N_t+M_td(N_t)+d(M_t)d(N_t)
\end{equation*}
for adapted quantum stochastic processes $M_t$, $N_t$ where
the quantum It\^o correction term (last term) is calculated
using the multiplication table
\begin{equation}\label{eq mult}\begin{array}{l}
(dt)^2=0,\quad dtdA_t=0=dtdA^*_t, \\dA_t^*dA_t=0,\quad dA_tdA_t^*=dt.
\end{array}\end{equation}

\subsection{Quantum Langevin Equations and Non-demolition Measurements}

From the quantum It\^o
formula applied to $X_t=U_t^*(X\otimes I)U_t$
and the quantum It\^o multiplication table (\ref{eq mult}),
we obtain the quantum Langevin equation
\begin{equation}\label{eq flowX}
dX_t=\LC_t[X_t]\otimes dt+[X_t,L_t]\otimes
dA^*_t-[X_t,L^*_t]\otimes dA_t.
\end{equation}
Here $\LC_t[X_t]=j_t(\LC[X])$
is the time evolved Lindblad (or Gorini-Kossakovski-Sudarshan)
generator \cite{GKS76,Lin76}
\begin{equation}\label{eq Lindblad}
\LC[X]=\frac{\textrm{i}}{\hbar}[H,X]+
\frac{1}{2}(L^*[X,L]+[L^*,X]L)
\end{equation}
for the semigroup of completely positive maps
describing the dissipative evolution in the Markovian
limit.
The dual $\LC^*$ of this map describes the
unconditional dissipative evolution of states
\begin{equation}\label{eq mastereq}
\frac{d}{dt}\rho^t=-\frac{\textrm{i}}{\hbar}[H,\rho^t]+\frac{1}{2}(L[\rho^t,L^*]+[L,\rho^t]L^*)
\end{equation}
called the
\emph{master equation} which is the quantum analogue
of the Focker-Plank equation.
A time continuous measurement of the
field quadrature
$W_t$
in the output
channel represents an indirect
measurement
of the evolved generalized coordinate
$L_t+L_t^*\vdash\AC_t$ as can be seen from
the quantum It\^o formula applied
to the output operators $Y_t=U_t^*(I\otimes
W_t)U_t$:
\begin{equation}\label{eq flowY}
dY_t=(L_t+L^*_t)\otimes dt+I\otimes dW_t.
\end{equation}
Note that the output process $Y_t$ is directly observable as it is
a commutative family of self-adjoint operators $\{Y_s\}_{s\le t}$
unitary equivalent to the family $\{W_s\}_{s\le t}$ for each $t$.
This simply follows from the following lemma which was first
observed by Belavkin in \cite{Bel79},\cite{Bel80}.
\begin{lem} The input and output
operators satisfy the quantum non-demolition (QND) condition
\begin{equation}\label{eq QND}
[X_t,Y_s]=0 \quad [Y_t,Y_s]=0 \quad \forall 0\le s\le t
\end{equation}
\end{lem}
\begin{proof}
Let $t=s+r, r>0$, then from the
cocycle identity we get
\begin{displaymath}\begin{array}{l}
U_{s+r}^*(I\otimes W_s)U_{s+r}= \\
\qquad U_s^*(S_{-s}U_rS_s)^*(I\otimes W_s)(S_{-s}U_rS_s)U_s= \\
\qquad\qquad U^*_s(I\otimes W_s)U_s=Y_s
\end{array}\end{displaymath}
where the last step uses the commutativity of
$S_{-s}U_rS_s\vdash \AC\otimes\WC_s^{s+r}$
and $W_s\vdash\WC_0^s$.
So $[X_t,Y_s]=U_t^*[X,W_s]U_t=0$
and $[Y_t,Y_s]=U_t^*[W_t,W_s]U_t=0$ follows from
the tensor independence of $X$, $W_s$ and
$W_t$ for all $s\ne t$.
\end{proof}

\subsection{Quantum Filtering}
Classically, filtering equations are used when we need to estimate
the value of dynamical variables about which we have incomplete
knowledge due to an indirect observation.  For example, the Kalman-Bucy filter
\cite{Kal60},\cite{KaB61} gives a continuous least-squares
estimator for a Gaussian classical random variable with linear
dynamics when we only have access to a correlated, noisy output
signal. Since closed quantum systems are fundamentally
unobservable unless they are open, e.g.\ disturbed by quantum
noise processes (c.f. (\ref{eq flowX}),(\ref{eq flowY})),
filtering of quantum noise plays an important role in quantum measurement.
Belavkin was the first to realize that an optimal estimation
without further disturbance is possible in the Markovian limit
and is based on an output nondemolition measurement
\cite{Bel79},\cite{Bel80},\cite{Bel88}. He constructed the quantum
filtering equation which describes the evolution of the optimal estimate
given by the density
matrix conditioned on a classical output of the
noisy quantum channel. This is used to estimate arbitrary input
operators $X_t\vdash\AC_t$ which are driven by environmental
quantum noises.  The previous lemma shows that the expectation of
$X_t$ is not disturbed when we measure $Y_s$ for $0\le s\le t$.
This is necessary for the existence of a well defined conditional
expectation of $X_t$ with respect to past measurement results of
$Y_s$.

Let $\CC_s^t:=\{Y_s^t\}''$ be the abelian von Neumann algebra
generated by the
output operators $Y_s^t:=\{Y_r|s\le r\le t\}$ (or their spectral
projectors in the case of unbounded $Y_r$). Also let
$\AC_s^t=\{X_r|s\le r\le t\}''$ denote the von Neumann algebra
generated by the system operators $X_r\vdash \AC_r$. From the QND
condition, $\CC_0^t$ lies in the center of (i.e.\ it is a subalgebra
commuting with the whole of)
$\BC_t^T\subset\AC\otimes\WC_0^T$, where
$\BC_t^T:=\AC_t^T\vee\CC_0^T$ is the smallest von Neumann algebra
containing $\AC_t^T$ and $\CC_0^T$ as subalgebras for $0\le t\le
T$. This gives the necessary conditions for the existence of a
conditional expectation \cite{Tak71}, defined as a linear,
normcontractive projection $E_0^T:\BC_t^T\to\CC_0^t$.

The conditional expectation $\mathbb{E}[X_t|Y_0^t]:=E_0^t[X_t]$
gives the least squares estimator $\hat{X}_t$ of
an operator $X_t\vdash\AC_t$ conditional on the output operators
$Y_0^t$ and so is equivalent to a classical random variable on the space of
measurement trajectories $\Omega_0^t:=\{\omega_s|0\le s\le t$ s.t.
$\omega_s$ is an eigenvalue of $Y_s\}$.  This conditional
expectation is most conveniently written in the Schr\"odinger
picture $E_0^t[X_t]=\langle \rho_\bullet^t,X\rangle$ for the solution
$\rho^t_\bullet$ to the classical stochastic nonlinear differential equation
\cite{Bel92b}
\begin{equation}\label{eq norm}
d\rho^t_\bullet=\LC^*[\rho^t_\bullet]dt+\sigma(\rho^t_\bullet)(dY_t-
\langle\rho^t_\bullet,L+L^*\rangle dt)
\end{equation}
often called the Belavkin quantum filtering equation, where
\begin{displaymath}
\sigma(\rho^t_\bullet)=\rho^t_\bullet L^*+L\rho^t_\bullet-
\langle\rho^t_\bullet,L^*+L\rangle
\rho^t_\bullet
\end{displaymath}
is the nonlinear \emph{fluctuation coefficient}.

We can generalize the filtering equation to the case where we
couple the open quantum system to $d$ independent measurement
channels.  If we assume no scattering between the channels, then
the family of unitary operators $\{U_t\}_{t\in\mathbb{R}_+}$
describing the evolution in the interaction picture
$U_t:\HC\otimes \FC_{[0,t)}^{\otimes d} \to \HC\otimes
\FC_{[0,t)}^{\otimes d}$ satisfy
\begin{displaymath}
dU_t+KU_t\otimes dt=\sum_{i=1}^d[L_iU_t\otimes dA_{i,t}^*-L_i^*U_t\otimes
dA_{i,t}]
\end{displaymath} where $L_i$ describes the coupling to the
$i$th channel and
$K=\frac{\textrm{i}}{\hbar}H+\frac{1}{2}\sum_{i=1}^dL^*_iL_i$.
Throughout this paper we reserve the Roman character $\textrm{i}$ to denote
the imaginary unit $\textrm{i}:=\sqrt{-1}$, whereas
italic $i$ is freely used as an index.
Note that we
have tensor independence of the annihilation increments
$dA_{i,t},dA_{j,t}$ for $i\ne j$, so the quantum vacuum noises
commute for different channels. The Belavkin filtering equation
for a simultaneous diffusive measurement of
$Y_{i,t}=U_t^*(I\otimes W_{i,t})U_t$ gives
\begin{equation}\label{eq filtd}
d\rho^t_\bullet=\LC^*[\rho^t_\bullet]dt+\sum_{i=1}^d\sigma_i(\rho^t_\bullet)(dY_{i,t}-
\langle\rho^t_\bullet,L_i+L_i^*\rangle dt)
\end{equation}
for $W_{i,t}=(A_{i,t}+A^*_{i,t})$.

\section{Optimal Quantum Control}

We now couple the system to a control field.  If we assume no
scattering between the measurement and control fields and
assume a weak coupling such that information is not
lost into the control field, then this effectively
replaces the Hamiltonian $H$ of the system with a controlled
Hamiltonian $H(u_s)$ for admissible real valued control functions
$u_s\in\mathbb{R}$ say,
at time $s$. This Hamiltonian generates the controlled unitaries
$U_t(u_0^t)$  giving the controlled flow
\begin{displaymath}
j_t(u_0^t)[X]:=U_t^*(u_0^t)(X\otimes I)U_t(u_0^t)
\end{displaymath}
where $u_0^t:=\{u_s|0\le s<t\}$ is
the control process over the interval $[0,t)$.
The controlled posterior density operator
$\rho^t_\bullet(u_0^t)$ can then be obtained
from (\ref{eq filtd})
with the controlled Hamiltonian $H(u_t)$ which
appears in the controlled Lindblad term $\LC(u_t)$.

In classical control, we can allow complete observability of the
controllable system, so that feedback controls are determined by
the system variables $x_t\to u_t(x_t)$. However, in
quantum systems, we do not have the point states $x_t$ due to
joint non observability of the system operators $X_t$, so the
stochastic feedback controls should be given by a function of the
stochastic output process $Y_0^t$ which is associated
with a classical random variable $u_t(\cdot)$ on $\Omega_0^t$.
I.e.\ the measurement trajectory
is fed into the control $\omega_0^t\mapsto u_t(\omega_0^t)$.
Thus the feedback controlled flow is a map
$j_t(u_0^t(Y_0^t))$ from $\AC$ to $\AC_t\vee\CC_0^t$.

The optimality of control is judged by the expected cost
associated to the admissible control process $u_0^T$
for the finite duration
$T$ of the experiment.  Admissible control strategies
are defined as those $u_0^T$ for which the
operator valued cost integral
\begin{equation}\label{eq cost}
J(u_0^T)=\int_0^Tj_s(u_0^s)[C(u_s)]ds+j_T(u_0^T)[S]
\end{equation}
exists in the strong operator topology
for self adjoint positive operators $C(u_s),S\vdash\AC$
giving the expected cost by the expectation
\begin{equation}
\langle\rho\otimes\phi,J(u_0^T)\rangle.
\end{equation}

An optimal feedback control strategy
$u_0^{T*}(\cdot)$ for nondemolition measurements
of the output operators $Y_0^T$
is one which minimizes the expected
posterior cost-to-go
\begin{displaymath}
\langle \rho\otimes\phi,J(u_0^{T*}(\cdot))\rangle=\min_{u_0^T(\cdot)\in U_0^T(\cdot)}
\langle \rho\otimes\phi,J(u_0^T(\cdot))\rangle
\end{displaymath}
where $U^T_0(\cdot)$ is the space of admissible stochastic control
strategies $u_0^T(\cdot)$.
This dynamical optimization problem is
considerably simplified by the following Lemma
first observed by Bellman.
\begin{lem}[Principle of Optimality]
If $u_0^{T*}(\cdot)$ is an optimal strategy for
the cost function (\ref{eq cost}) given the
initial state $\rho\otimes\phi$, then its restriction
$u_t^{T*}(\cdot)$ to the interval $[t,T)$ is
optimal for the \emph{cost-to-go}
\begin{equation}\label{eq costtogo}\begin{array}{l}
J_t(u_t^T(\cdot))=\int_t^Tj_s(u_t^s(\cdot))[C(u_s(\cdot))]ds\\
\qquad\qquad\quad+j_T(u_t^T(\cdot))[S]
\end{array}\end{equation}
given the state $\rho_\bullet^t(u_0^t(\cdot))$ at time $t$.
\end{lem}

We can now reduce the dynamics to the observable
output algebra and rewrite the expectation
as a conditional one
\[
\langle \rho\otimes\phi,J_t(u_t^T(Y_0^t))\rangle=
\langle \phi_0^t,E_0^t[J_t(u_t^T(Y_0^t))]\rangle
\]
where $E_0^t:\BC_t^T\to\CC_0^t$ is the conditional expectation on
$\BC_t^T=\AC_t^T\vee\CC_0^T$ which defines the feedback controlled
posterior density operator by $\langle\rho^t_\bullet(u_0^t(\cdot)),X\rangle=E_0^t\circ
j_t(u_0^t(\cdot))[X]$ at time $t$.

\begin{thm}
The posterior cost-to-go from state $\rho$ at time $t$ satisfies
\begin{equation}
E_0^t[J_t(u_t^T(\cdot))]=
\mathbb{E}_0^t[\textsf{J}(t,u_t^T(\cdot),\rho)]
\end{equation}
where
\begin{eqnarray*}
\textsf{J}(t,u_t^T(\cdot),\rho)=&
\int_t^T\langle\rho^s_\bullet(u_t^s(\cdot)),C(u_s(\cdot))\rangle
ds \\ &+\langle\rho^T(u_t^T(\cdot)),S\rangle
\end{eqnarray*}
is a random variable on $\Omega_0^T$ and $\rho^s_\bullet(u_t^s(\cdot))$ is
the solution to the controlled filtering equation for $s\ge t$
with $\rho=\rho^t_\bullet(u_0^t(\cdot))$.
\end{thm}
\begin{proof}
The 'quantum' conditional expectation $E_0^t$ acting on future
operators gives
\begin{displaymath}
E_0^t\circ j_s(u_t^s(\cdot))[X]=
\mathbb{E}_0^t[\langle\rho^s_\bullet(u_t^s(\cdot)),X\rangle]
\end{displaymath}
for $X\vdash\AC$, where $\mathbb{E}_0^t:\CC_0^T\to\CC_0^t$ is the
'classical'
conditional expectation on $\CC_0^T$ satisfying the tower property
$\mathbb{E}_0^t\circ\mathbb{E}_0^s=\mathbb{E}_0^t$ for $t\le s\le
T$.
\end{proof}

Let us denote the minimum posterior cost-to-go
\begin{equation}
\textsf{S}(t,\rho):=\min_{u_t^T(\cdot)\in
U_t^T(\cdot)}\mathbb{E}_0^t[\textsf{J}(t,u_t^T(\cdot),\rho)].
\end{equation}
\begin{thm} The minimum posterior
cost-to-go satisfies the Bellman equation
\begin{displaymath}
\frac{\partial}{\partial t}\textsf{S}(t,\rho)+
\frac{1}{2}\sum_{i=1}^d\langle\sigma_i(\rho)\otimes\sigma_i(\rho),
(\delta\otimes\delta)\textsf{S}(t,\rho)\rangle \\
\end{displaymath}
\begin{equation}\label{eq Bellman}
+\min_{u_t(\cdot)}\left\{\langle\rho,C(u_t(\cdot))+
\LC(u_t(\cdot))[\delta\textsf{S}(t,\rho)]\rangle\right\}=0
\end{equation}
where $\delta\textsf{S}(t,\rho)\vdash\AC$ denotes the derivation
of $\textsf{S}(t,\rho)$ with respect to $\rho$ and $\sigma_i(\rho)$
is the non-linear fluctuation coefficient in the filtering equation
(\ref{eq filtd}).
\end{thm}
\begin{proof}
From the definition of $\textsf{S}(t,\rho)$ and
$\textsf{J}(t,u_t^T(\cdot),\rho)$, we have
\begin{displaymath}
\textsf{S}(t,\rho^t)=\min_{u_t^T(\cdot)}
\mathbb{E}_0^t\left\{
\begin{array}{l}
\int_t^{t+\epsilon}\langle\rho^s_\bullet(u_t^s(\cdot)),C(u_s(\cdot))\rangle ds \\
+\textsf{J}(t+\epsilon,u_{t+\epsilon}^T(\cdot),\rho^{t+\epsilon})
\end{array}\right\}
\end{displaymath}
So when $\epsilon\to dt$ becomes sufficiently small, we approximate this by
\begin{equation}\label{eq prebell}
\textsf{S}(t,\rho^t)=\min_{u_t(\cdot)}
\mathbb{E}_0^t\left\{
\begin{array}{l}
\langle\rho^t,C(u_t(\cdot))\rangle dt \\
+\textsf{S}(t+dt,\rho^{t+dt})
\end{array}\right\}
\end{equation}
where we use the tower property of the classical conditional expectation.
Assuming that $\textsf{S}(t,\rho^t)$ is sufficiently differentiable,
we use the Taylor expansion
\begin{equation*}\begin{array}{l}
\textsf{S}(t+dt,\rho^{t+dt})= \\
\quad\textsf{S}(t,\rho^t)+
\frac{\partial}{\partial t}\textsf{S}(t,\rho^t)dt+
\langle d\rho^t,\delta\textsf{S}(t,\rho^t)\rangle+ \\
\quad\frac{1}{2}\sum_{i=1}^d\langle\sigma_i(\rho^t)\otimes\sigma_i(\rho^t),
(\delta\otimes\delta)\textsf{S}(t,\rho^t)\rangle dt
\end{array}\end{equation*}
where
$\delta\textsf{S}(t,\rho):=\frac{\delta}{\delta\rho}\textsf{S}(t,\rho)$
denotes the derivation of $\textsf{S}(t,\rho)$ with respect to
$\rho$. Using this expansion in (\ref{eq prebell}) gives the Bellman
equation (\ref{eq Bellman}) when we observe that $\textsf{S}(t,\rho)+
\frac{\partial}{\partial t}\textsf{S}(t,\rho)$ does not depend on
$u_t$ and $\mathbb{E}_0^t[d\tilde{Y}_{i,t}]=0$ for the innovation
process $d\tilde{Y}_{i,t}=dY_{i,t}- \langle\rho^t,L_i+L_i^*\rangle dt$.
\end{proof}

\section{Application of Results to a Linear Quantum Dynamical System}

We illustrate the ideas of quantum filtering and
control described above by application to the
multidimensional quantum LQG control problem.
LQG control is well studied in classical control
theory and we shall see many similarities between
quantum and classical LQG control theory.

\subsection{Quantum Filtering of Linear, Gaussian Dynamics}

Let $\boldsymbol{X}$ be the
phase space vector of self adjoint operators
$X^i$, $i=1,...,m$
satisfying the canonical commutation relations (CCRs)
\begin{displaymath}
[X^i,X^j]=X^iX^j-X^jX^i=\textrm{i}\hbar J^{ij}I
\end{displaymath}
for $i,j=1,...m$ where $I$ is the identity operator on $\HC$.
The CCRs can be written in vector form as
\begin{displaymath}
[\boldsymbol{X},\boldsymbol{X}^\top]:=\boldsymbol{X}\boldsymbol{X}^\top
-(\boldsymbol{X}\boldsymbol{X}^\top)^\top=\textrm{i}\hbar\mathbf{J}I
\end{displaymath}
where $\boldsymbol{X}^\top=(X^1,...,X^{m})$
is the row vector transpose of $\boldsymbol{X}$ and
$\mathbf{J}=(J^{ij})$ is an anti-symmetric real valued matrix
which is assumed to be nondegenerate for an even $m=2d$ say.
We couple the open quantum
system to $d$ measurement channels via the
operator vector
$\boldsymbol{L}=\mathbf{\Lambda}\boldsymbol{X}$,
where $\mathbf{\Lambda}$ is a $d\times m$ matrix
of complex-valued coefficients. Let us place it
in a controllable potential
which is described by the Hamiltonian
\begin{equation}\label{eq Hamiltonian}
H(\boldsymbol{u}_t)=\frac{1}{2}\boldsymbol{X}^\top\mathbf{R}\boldsymbol{X}
+\boldsymbol{X}^\top\mathbf{K}\boldsymbol{u}_t+\boldsymbol{u}_t^\top\mathbf{K}^\dagger\boldsymbol{X}
\end{equation}
for real vector valued control parameters
$\boldsymbol{u}_t\in\mathbb{R}^d$, where $\mathbf{R}$ is a real symmetric $m\times m$ matrix
and $\mathbf{K}$ is a complex $m\times d$ matrix.
We shall use $\mathbf{\Lambda}^*$ to denote complex conjugation
$(\mathbf{\Lambda}^*)_{ij}=\Lambda_{ij}^*$ and
$\mathbf{\Lambda}^\dagger=(\mathbf{\Lambda}^*)^\top$ the Hermitian
conjugate.

These definitions allow us to calculate the
components of the controlled Lindblad generator
from (\ref{eq Lindblad}) with the controlled
Hamiltonian (\ref{eq Hamiltonian})
which we write here in vector form
\begin{displaymath}
\LC(\boldsymbol{u}_t)[\boldsymbol{X}]=
\mathbf{J}(\mathbf{R}+\hbar\Im(\mathbf{\Lambda}^\dagger\mathbf{\Lambda}))\boldsymbol{X}
+\mathbf{J(K+K^*)}\boldsymbol{u}_t
\end{displaymath}
omitting the identity $I$ for notational convenience
where $2i\Im(\mathbf{\Lambda}^\dagger\mathbf{\Lambda})=
\mathbf{\Lambda}^\dagger\mathbf{\Lambda}-
\mathbf{\Lambda}^\top\mathbf{\Lambda}^*$.
So from (\ref{eq flowX}) and (\ref{eq flowY}) we obtain
the following quantum linear Langevin vector equation
\begin{equation}\label{eq Langevin}
d\boldsymbol{X}_t=(\mathbf{A}\boldsymbol{X}_t+\mathbf{B}\boldsymbol{u}_t)dt+d\boldsymbol{V}_t
\end{equation}
and linear output equation
\begin{equation}
d\boldsymbol{Y}_t=\mathbf{C}\boldsymbol{X}_tdt+d\boldsymbol{W}_t
\end{equation}
where
$\mathbf{A}:=\mathbf{J}(\mathbf{R}+\hbar\Im(\mathbf{\Lambda}^\dagger\mathbf{\Lambda}))$,
$\mathbf{B}:=\mathbf{J(K+K^*)}$,
$\mathbf{C}:=\mathbf{\Lambda+\Lambda}^*$.  The quantum
noise increments are given by vectors
\begin{eqnarray*}
d\boldsymbol{V}_t&=&i\hbar\mathbf{J}(\mathbf{\Lambda}^\top
d\boldsymbol{A}^*_t-\mathbf{\Lambda}^\dagger d\boldsymbol{A}_t) \\
d\boldsymbol{W}_t&=&d\boldsymbol{A}_t+d\boldsymbol{A}_t^*
\end{eqnarray*}
for $(\boldsymbol{A}_t)_i=A_{i,t}$ the
annihilation operator on the $i$th coupled
independent measurement channel.

Let us
denote the initial mean $\bar{\boldsymbol{X}}$ of the phase space operator vector
by the component wise expectation
$(\bar{\boldsymbol{X}})^i=\bar{X}^i=\langle\rho,X^i\rangle$ and symmetric
covariance
\begin{displaymath}
\Sigma^{ij}:=\frac{1}{2}\langle\rho,X^iX^j+X^jX^i\rangle-
\bar{X}^i\bar{X}^j.
\end{displaymath}
which is given by a real positive definite matrix $\mathbf{\Sigma}=(\Sigma^{ij})$
satisfying the Heisenberg uncertainty principle
\begin{equation}\label{eq HUP}
\mathbf{\Sigma}\ge\pm\frac{\textrm{i}\hbar}{2}\mathbf{J}
\end{equation}

The filtering equation (\ref{eq filtd}) preserves the Gaussian nature of the
posterior state \cite{Bel92a}, so the posterior mean
$(\hat{\boldsymbol{X}_t})^i=\hat{X}_t^i=\langle\rho^t_\bullet,X^i\rangle$
and symmetric error covariances
\[
\Sigma^{ij}_t:=\frac{1}{2}
\langle\rho^t_\bullet,X^iX^j+X^jX^i\rangle-
\hat{X}^i_t\hat{X}^j_t
\]
form a set of sufficient coordinates for
the quantum LQG system and agree with the initial
mean and covariance for $\rho_\bullet^0=\rho$.
Using (\ref{eq filtd}), the posterior expectation of
$\boldsymbol{X}_t$
for non demolition measurement of the output operators $\boldsymbol{Y}_t$ is
given in vector form
\begin{eqnarray}\label{eq LQGfilt}
d\hat{\boldsymbol{X}}_t&=&(\mathbf{A}\hat{\boldsymbol{X}}_t+\mathbf{B}\boldsymbol{u}_t)dt
+\tilde{\mathbf{K}}_td\tilde{\boldsymbol{Y}}_t \\
\tilde{\mathbf{K}}_t&=&\mathbf{\Sigma}_t\mathbf{C}^\top+\mathbf{M}
\end{eqnarray} where
$d\tilde{\boldsymbol{Y}}_t=d\boldsymbol{Y}_t-\mathbf{C}\hat{\boldsymbol{X}}_tdt$
is the innovating martingale which describes the information gain
from measurement of the output vector operator $\boldsymbol{Y}_t$.

The symmetric error covariance
$\mathbf{\Sigma}_t$ satisfies the matrix Ricatti equation
\begin{equation}\label{eq FiltRicatti}\begin{array}{rl}
\frac{d}{dt}\mathbf{\Sigma}_t=&\mathbf{A\Sigma}_t+\mathbf{\Sigma}_t\mathbf{A}^\top+\mathbf{N} \\
&-(\mathbf{\Sigma}_t\mathbf{C}^\top+\mathbf{M})(\mathbf{\Sigma}_t\mathbf{C}^\top+\mathbf{M})^\top\\
\mathbf{\Sigma}_0=&\mathbf{\Sigma}
\end{array}\end{equation}
where
\[
\mathbf{N}=\frac{1}{2}\hbar^2\mathbf{J}(\mathbf{\Lambda}^\dagger\mathbf{\Lambda}+
\mathbf{\Lambda}^\top\mathbf{\Lambda}^*)\mathbf{J}^\top
\]
is the intensity (symmetric covariance) matrix of the quantum noise
increment
$d\boldsymbol{V}_t$ and
\[
\mathbf{M}=\frac{i}{2}\hbar\mathbf{J}(\mathbf{\Lambda}^\top-\mathbf{\Lambda}^\dagger)
\]
is the
covariance matrix of the noise increments $d\boldsymbol{V}_t$ and
$d\boldsymbol{W}_t$.

\subsection{Quantum LQG Control}
We aim to control the phase space operator
whilst constraining the amplitude of the
controlling force for energy considerations.
Thus, our control objectives and restraints
can be described by the
operator valued risk (\ref{eq cost})
with quadratic parameters
\[
C(\boldsymbol{u}_s)=\boldsymbol{X}^\top\mathbf{F}\boldsymbol{X}+
\boldsymbol{X}^\top\mathbf{G}^\top\boldsymbol{u}_s+
\boldsymbol{u}_s^\top\mathbf{G}\boldsymbol{X}+
\boldsymbol{u}_s^\top\boldsymbol{u}_s
\]
and $S=\boldsymbol{X}^\top\mathbf{\Omega}\boldsymbol{X}$
for positive real symmetric $m\times m$ matrices $\mathbf{\Omega,F}$
and a real $d\times m$ matrix $G$.

Since $\hat{\boldsymbol{X}}$ and $\mathbf{\Sigma}$ form a set of
sufficient coordinates,
they describe the full probability distribution
given by $\rho$, so we may consider the derivation of
$\textsf{S}(t,\rho)$ as partial derivatives of
$\textsf{S}(t,\hat{\boldsymbol{X}},\mathbf{\Sigma})$.  So from (\ref{eq
Langevin}) and the Gaussian nature of the system, we obtain
\begin{displaymath}\begin{array}{l}
\langle\rho,\LC(\boldsymbol{u}_t)[\delta\textsf{S}(t,\hat{\boldsymbol{X}},\mathbf{\Sigma})]\rangle= \\
\quad\frac{1}{2}(\mathbf{A}\hat{\boldsymbol{X}}+\mathbf{B}\boldsymbol{u}_t)^\top\nabla_{\hat{\boldsymbol{X}}}\textsf{S}
+\frac{1}{2}\nabla_{\hat{\boldsymbol{X}}}\textsf{S}^\top(\mathbf{A}\hat{\boldsymbol{X}}+\mathbf{B}\boldsymbol{u}_t) \\
\quad+\left(\mathbf{A\Sigma+\Sigma A}^\top+\mathbf{N},\nabla_{\mathbf{\Sigma}}\textsf{S}\right) \\
\end{array}\end{displaymath}
\begin{displaymath}\begin{array}{l}
\sum_{j=1}^d\langle\sigma_j(\rho)\otimes\sigma_j(\rho),(\delta\otimes\delta)\textsf{S}
(t,\hat{\boldsymbol{X}},\mathbf{\Sigma})\rangle=\qquad\\
\quad\left((\mathbf{\Sigma C}^\top+\mathbf{M)(\Sigma
C}^\top+\mathbf{M})^\top,\nabla^2_{\hat{\boldsymbol{X}}}\textsf{S}-2\nabla_{\mathbf{\Sigma}}\textsf{S}\right)
\end{array}\end{displaymath}
where
$(\mathbf{D},\mathbf{E}):=\textrm{Tr}[\mathbf{D}^\top\mathbf{E}]$ is the
Hilbert-Schmidt inner product on the vector space of
complex-valued $m\times m$ matrices.
We denote the partial derivatives by $(\nabla_{\hat{\boldsymbol{X}}}\textsf{S})_i=\frac{\partial}{\partial
\hat{\boldsymbol{X}}^i}\textsf{S}(t,\hat{\boldsymbol{X}},\mathbf{\Sigma})$,
$(\nabla_{\mathbf{\Sigma}}\textsf{S})_{ij}=\frac{\partial}{\partial
\mathbf{\Sigma}_{ij}}S(t,\hat{\boldsymbol{X}},\mathbf{\Sigma})$ and
$(\nabla^2_{\hat{\boldsymbol{X}}}\textsf{S})_{ij}=\frac{\partial}{\partial
\hat{\boldsymbol{X}}^i}\frac{\partial}{\partial
\hat{\boldsymbol{X}}^j}\textsf{S}(t,\hat{\boldsymbol{X}},\mathbf{\Sigma})$.
Inserting into the Bellman equation (\ref{eq Bellman}) and minimizing
gives
$\boldsymbol{u}_t=-(\frac{1}{2}\mathbf{B}^\top\nabla_{\hat{\boldsymbol{X}}}\textsf{S}
+\mathbf{G}\hat{\boldsymbol{X}})$ where
$\textsf{S}(t,\hat{\boldsymbol{X}},\mathbf{\Sigma})$ now satisfies the
nonlinear partial differential equation
\begin{equation}\label{eq HJB}\begin{array}{l}
-\frac{\partial}{\partial
t}\textsf{S}(t,\hat{\boldsymbol{X}},\mathbf{\Sigma})= \\
\quad\frac{1}{2}(\hat{\boldsymbol{X}}^\top\mathbf{A}^\top\nabla_{\hat{\boldsymbol{X}}}\textsf{S}+
\nabla_{\hat{\boldsymbol{X}}}\textsf{S}^\top\mathbf{A}\hat{\boldsymbol{X}})
+\hat{\boldsymbol{X}}^\top\mathbf{F}\hat{\boldsymbol{X}}\\
\quad+\left(\mathbf{A\Sigma+\Sigma A}^\top+\mathbf{N},\nabla_{\mathbf{\Sigma}}\textsf{S}\right)
+\left(\mathbf{\Sigma},\mathbf{F}\right) \\
\quad-(\frac{1}{2}\mathbf{B}^\top\nabla_{\hat{\boldsymbol{X}}}\textsf{S}
+\mathbf{G}\hat{\boldsymbol{X}})^\top(\frac{1}{2}\mathbf{B}^\top\nabla_{\hat{\boldsymbol{X}}}\textsf{S}
+\mathbf{G}\hat{\boldsymbol{X}}) \\
\quad+\left((\mathbf{\Sigma C}^\top+\mathbf{M)(\Sigma
C}^\top+\mathbf{M})^\top,\frac{1}{2}\nabla^2_{\hat{\boldsymbol{X}}}\textsf{S}-\nabla_{\mathbf{\Sigma}}\textsf{S}\right)
\end{array}\end{equation}
which is called the Hamilton-Jacobi-Bellman (HJB) equation for this example.

It is well known from classical control theory that LQG control
gives a posterior cost-to-go which is quadratic in the posterior
mean.  So we use the ansatz
\begin{displaymath}
\textsf{S}(t,\hat{\boldsymbol{X}},\mathbf{\Sigma})=
\hat{\boldsymbol{X}}^\top\mathbf{\Omega}_t\hat{\boldsymbol{X}}
+\langle\mathbf{\Omega}_t,\mathbf{\Sigma}\rangle +\alpha_t
\end{displaymath}
in the HJB equation (\ref{eq HJB}).
This gives the optimal feedback control strategy
\begin{eqnarray}\label{eq optcont}
\boldsymbol{u}_t&=&-\tilde{\mathbf{L}}_t\hat{\boldsymbol{X}}_t \\ \label{eq L}
\tilde{\mathbf{L}}_t&=&\mathbf{B}^\top\mathbf{\Omega}_t+\mathbf{G}
\end{eqnarray}
which is linear in the solution to the filtering equation
$\hat{\boldsymbol{X}}_t$ at time $t$ where $\mathbf{\Omega}_t$ satisfies
the backwards matrix Ricatti equation
\begin{equation}\label{eq ContRicatti}\begin{array}{rl}
-\frac{d}{dt}\mathbf{\Omega}_t=&\mathbf{\Omega}_t\mathbf{A}+\mathbf{A}^\top\mathbf{\Omega}_t+\mathbf{F}\\
&-(\mathbf{B}^\top\mathbf{\Omega}_t+\mathbf{G})^\top(\mathbf{B}^\top\mathbf{\Omega}_t+\mathbf{G})\\
\mathbf{\Omega}_T=&\mathbf{\Omega}
\end{array}\end{equation}
and $\alpha_t$ satisfies
\begin{equation}\begin{array}{rl}
-\frac{d}{dt}\alpha_t=&\left((\mathbf{B}^\top\mathbf{\Omega+G})^\top(\mathbf{B}^\top\mathbf{\Omega+G}),\mathbf{\Sigma}_t\right) \\
&+\left(\mathbf{\Omega}_t,\mathbf{N}\right) \\
\alpha_T=&0.
\end{array}\end{equation}
From this we obtain the total minimal cost
\begin{equation}\label{eq totalcost}\begin{array}{l}
\textsf{S}(0,\bar{\boldsymbol{X}},\mathbf{\Sigma})= \\
\quad\bar{\boldsymbol{X}}^\top\mathbf{\Omega}_0\bar{\boldsymbol{X}}
+\textrm{Tr}[\mathbf{\Omega}_0\mathbf{\Sigma}]+\int_0^T\textrm{Tr}[\mathbf{\Omega}_t\mathbf{N}]dt  \\
\quad+\int_0^T\textrm{Tr}[(\mathbf{B}^\top\mathbf{\Omega+G})^\top(\mathbf{B}^\top\mathbf{\Omega+G})\mathbf{\Sigma}_t]dt
\end{array}\end{equation}
where $\mathbf{\Omega}_0$ is the solution to (\ref{eq ContRicatti}) at time $t=0$.

\subsection{Duality}

The example of the quantum LQG control
problem is important since it is one
of the few exactly solvable control
problems and emphasizes the
similarities between the two components
of optimal quantum feedback control,
namely
quantum filtering and optimal control.
The duality between the solutions of
filtering (\ref{eq LQGfilt})-(\ref{eq FiltRicatti})
and control (\ref{eq optcont})-(\ref{eq ContRicatti})
is summarized in the duality table
\begin{equation}\label{eq duality}
\begin{array}{c|c|c|c|c|c|c}
\textrm{Filtering} & \mathbf{\Sigma}_t & \tilde{\mathbf{K}}_t &
\mathbf{A} & \mathbf{C} & \mathbf {N} & \mathbf {M} \\
\hline
\textrm{Control} & \mathbf{\Omega}_{T-t} & \tilde{\mathbf{L}}^\top_{T-t} &
\mathbf{A}^\top & \mathbf{B}^\top & \mathbf{F} & \mathbf{G}^\top
\end{array}
\end{equation}

which allows us to formulate and solve the dual control
problem given the filtering parameters.
The duality can be understood when we examine
the nature of each of the methods used.
Both methods involve the minimization of a
quadratic function for linear, Gaussian
systems, (i.e.\ the least squares error
for filtering and the quadratic cost
for control).  The time reversal in the
dual picture is explained by the forward (backward)
induction used in the dynamical minimization problem for
the filtering (control) problem.

\subsection{Optimal feedback control of continuously observed
quantum free particle}

We give a more physical interpretation of the above results
by application to an explicit example of LQG control where
the duality between filtering and control is preserved.  The example
of the complex Gaussian oscillator was given in \cite{Bel99}, however
we may now use the multidimensional quantum LQG control
solutions derived above for application on higher dimensional
systems which do not have such complex representation.  The
optimal control of a continuously observed quantum free particle
with quadratic cost is the simplest such example.

Let $\boldsymbol{X}^\top=(Q,P)$ be the phase space
vector operator consisting of the position $Q$ and
momentum $P$ operators of the free particle having
the initial expectations $\bar{Q}$ and $\bar{P}$
respectively.  Let us also denote the initial
dispersions by $\sigma_Q$ and $\sigma_P$ respectively
and the initial covariance of $Q$ and $P$ by
$\sigma_{QP}=\sigma_{PQ}$.
We can perform a continuous observation of the
particle by coupling the position operator
to the measurement channel $L=Q$ in which we measure
the classical Wiener process
$W_t=A_t+A_t^*$ and the particle is controlled
using the linear potential $V(u_t)=-u_tQ$
for $u_t\in\mathbb{R}$.
The Hamiltonian of this simple system is then given by
$H(u_t)=\frac{1}{2M}P^2-u_tQ$ where $M$ is the
mass of the particle and the corresponding
Langevin equations are
\begin{equation}\label{eq FPlang}
\frac{d}{dt}Q_t=\frac{1}{M}P_t
\qquad \frac{d}{dt}P_t=u_t+\dot{V}_t
\end{equation}
where $\dot{V}_t$ is the time derivative of the
Wiener process $V_t=\hbar W_t^{\pi/2}$ and
represents the system process noise
due to the interaction with the coupled
noise bath.
The operators $Y_t$ satisfy the
linear output equation
\begin{equation}
\frac{d}{dt}Y_t=2Q_t+\dot{W}_t
\end{equation}
which is perturbed by measurement noises
represented by the time derivative of the
Wiener process $W_t$.

The optimal estimates of the position
and momentum based on a non demolition
measurement of $Y_t$
are then given by the quantum Kalman
Bucy filter (\ref{eq LQGfilt})
\begin{eqnarray}
d\hat{Q}_t&=&\frac{1}{M}\hat{P}_tdt+2\sigma_{Q,t}d\tilde{Y}_t \\
d\hat{P}_t&=&u_tdt+2\sigma_{QP,t}d\tilde{Y}_t
\end{eqnarray}
where the innovation process $\tilde{Y}_t$ describes
the gain of information due to measurement of $Y_t$
given by
\[
\tilde{Y}_t=Y_t-2\hat{Q}_t.
\]
In practice, for a continuous observation, it is the
measurement current $I_t:=dY_t/dt$ which we observe
and so we write the filtering equations in the form
\begin{eqnarray}
\frac{d}{dt}\hat{Q}_t&=&\frac{1}{M}\hat{P}_t+2\sigma_{Q,t}(I_t-2\hat{Q}_t) \\
\frac{d}{dt}\hat{P}_t&=&u_t+2\sigma_{QP,t}(I_t-2\hat{Q}_t)
\end{eqnarray}
where the error covariances satisfy the
Ricatti equations
\begin{equation}\label{eq FPFiltRicatti}\begin{array}{lll}
\frac{d}{dt}\sigma_{Q,t}&=&\frac{2}{M}\sigma_{QP,t}-4(\sigma_{Q,t})^2 \\
\frac{d}{dt}\sigma_{QP,t}&=&\frac{1}{M}\sigma_{P,t}-4\sigma_{Q,t}\sigma_{QP,t} \\
\frac{d}{dt}\sigma_{P,t}&=&\hbar^2-4(\sigma_{QP,t})^2
\end{array}\end{equation}
with initial conditions
\[
\sigma_{Q,0}=\sigma_Q,\quad\sigma_{QP,0}=\sigma_{QP},\quad\sigma_{P,0}=\sigma_P.
\]

The Ricatti equations for the error covariance in the
filtered free particle dynamics have an exact solution
\cite{BeS92}, however we will simply comment on the
stationary solutions which are the solutions obtained
by setting the LHS of (\ref{eq FPFiltRicatti}) to zero, giving
the asymptotic behaviour of the posterior dispersions
for $t\to\infty$
\begin{equation}
\sigma_{Q,t}\to\frac{1}{2}\sqrt{\frac{\hbar}{M}}, \quad
\sigma_{P,t}\to \hbar\sqrt{\hbar M},\quad
\sigma_{PQ,t}\to\frac{\hbar}{2}.
\end{equation}
This proper treatment dispels the paradoxical quantum
Zeno effect which insists that a quantum state is
frozen in time by a continuous observation.  Instead we can describe
the continuous observation as an optimal estimation
with posterior dispersions tending to a finite limit
satisfying the Heisenberg uncertainty relation
\[
\Delta_{Q,t}\Delta_{P,t}=\sqrt{\sigma_{Q,t}\sigma_{P,t}}\to\hbar/\sqrt{2}\ge\hbar/2.
\]
In contrast, for the case without conditioning (where the measurement
results are ignored or averaged over) the Ricatti
equations for the dispersions become linear which have solutions tending
to infinity like $t^3$.  This is faster than the $t^2$ spreading
of the wavefunction due to the closed evolution described by Schr\"odinger's
equation as one would expect since the coupled noise bath only serves
to increase the dispersion.

The dual optimal control problem can be found by identifying
the corresponding dual matrices from the table (\ref{eq duality})
which give the quadratic control parameters
\begin{eqnarray*}
C(u_t)&=&\beta Q^2+u_t^2 \\
S&=&\omega_{Q}Q^2+\omega_{QP}(PQ+QP)+\omega_{P}P^2
\end{eqnarray*}
which for the linear Gaussian system (\ref{eq FPlang}) gives the
optimal control strategy
\begin{equation}
u_t=-2(\omega_{PQ,t}\hat{P}_t+\omega_{P,t}\hat{Q}_t)
\end{equation}
where the coefficients are
the solutions to the Ricatti equations
\begin{equation}\label{eq FPContRicatti}\begin{array}{lll}
-\frac{d}{dt}\omega_{P,t}&=&\frac{2}{M}\omega_{QP,t}-4(\omega_{P,t})^2 \\
-\frac{d}{dt}\omega_{QP,t}&=&\frac{1}{M}\omega_{Q,t}-4\omega_{P,t}\omega_{QP,t} \\
-\frac{d}{dt}\omega_{Q,t}&=&\beta-4(\omega_{QP,t})^2
\end{array}\end{equation}
with terminal solutions
\[
\omega_{P,0}=\omega_P,\quad\omega_{QP,0}=\omega_{QP},\quad\omega_{Q,0}=\omega_Q.
\]
Note that in this example, as well as identifying the dual matrices by
transposition and time reversal according to the duality table (\ref{eq duality}),
one must also interchange the coordinates $P\leftrightarrow Q$.  This is because
the matrix of coefficients $\mathbf{A}$ is non-symmetric and nilpotent, so it is
dual to its transpose only when we interchange the coordinates in the
dual picture.  Thus the optimal coefficients $\{\omega_{P,t},\omega_{QP,t},\omega_{Q,t}\}$
in the quadratic cost-to-go correspond to the minimal error covariances
$\{\sigma_{Q,T-t},\sigma_{QP,T-t},\sigma_{P,T-t}\}$ in the dual picture.

The minimal total cost for the experiment can be obtained from (\ref{eq totalcost})
by substitution of these solutions
\begin{equation}\label{eq FPtotalcost}\begin{array}{l}
\textsf{S}=\omega_{Q,0}(\bar{Q}^2+\sigma_Q)+2\omega_{QP,0}(\bar{Q}\bar{P}+\sigma_{QP})\\
\quad+\omega_{P,0}(\bar{P}^2+\sigma_P)+\int_0^T(\hbar^2\omega_{P,t}+\omega_{PQ,t}^2\sigma_{Q,t})dt\\
\quad+\int_0^T(\omega_{P,t}^2\sigma_{P,t}+2\omega_{QP,t}\omega_{P,t}\sigma_{PQ,t})dt
\end{array}\end{equation}

\section{Discussion}

We have shown that the optimal quantum feedback control problem
reduces to an optimal estimation problem followed by an optimal
control problem based on this optimal estimator.  The
optimal (least-squares) estimator for quantum random variables
(operators) given a classical nondemolition output
measurement process is the conditional expectation
which is given by the result of the filtering equation (\ref{eq filtd}).
The resulting optimal control problem is then defined
on the output of this filter, which reduces to a
classical control problem on the space of quantum states.
For cost functions that are linear in the state, the optimal
feedback control strategy is given by the solution
to the Bellman equation (\ref{eq Bellman}).

In the LQG example, the space of quantum states are restricted
to the class of Gaussian states so the probability
distribution is parameterized
by the mean and covariance of the generating operators.
However, due to non commutativity of these quantum operators
there are many different definitions of the covariance matrices.
For direct comparison to classical LQG control theory, we
choose the symmetric representation of the covariance matrices,
although unlike the classical case, the Heisenberg
uncertainty principle (\ref{eq HUP}) places a positive lower bound on
the covariances.  In particular, this prohibits
the common classical assumption of uncorrelated
process and measurement noise if the coupling to
the noise bath is complex.

\section*{Acknowledgements}
This work has been supported by the EPSRC under the programme
Mathfit (grant no RA2273).  VPB also acknowledges support from the
EC under the programme ATESIT (contract no IST-2000-29681).

\bibliography{F:/Documents/BibDB/Main}

\begin{thebibliography}{10}
\providecommand{\url}[1]{#1}
\csname url@rmstyle\endcsname
\providecommand{\newblock}{\relax}
\providecommand{\bibinfo}[2]{#2}
\providecommand\BIBentrySTDinterwordspacing{\spaceskip=0pt\relax}
\providecommand\BIBentryALTinterwordstretchfactor{4}
\providecommand\BIBentryALTinterwordspacing{\spaceskip=\fontdimen2\font plus
\BIBentryALTinterwordstretchfactor\fontdimen3\font minus
  \fontdimen4\font\relax}
\providecommand\BIBforeignlanguage[2]{{%
\expandafter\ifx\csname l@#1\endcsname\relax
\typeout{** WARNING: IEEEtran.bst: No hyphenation pattern has been}%
\typeout{** loaded for the language `#1'. Using the pattern for}%
\typeout{** the default language instead.}%
\else
\language=\csname l@#1\endcsname
\fi
#2}}

\bibitem{AASDM02}
M.~A. Armen, J.~K. Au, J.~K. Stockton, A.~C. Doherty, and H.~Mabuchi,
  ``Adaptive homodyne measurement of optical phase,'' \emph{Phys.~Rev.~Lett.},
  vol.~89, p. 133602, 2002.

\bibitem{GSDM03}
J.~M. Geremia, J.~K. Stockton, A.~C. Doherty, and H.~Mabuchi, ``Quantum
  {Kalman} filtering and the {Heisenberg} limit in atomic magnetometry,''
  \emph{Phys. Rev. Lett.}, vol.~91, p. 250801, 2003.

\bibitem{ADL02}
C.~Ahn, A.~C. Doherty, and A.~J. Landahl, ``Continuous quantum error correction
  via quantum feedback control,'' \emph{Phys.~Rev.~A}, vol.~65, p. 042301,
  2002.

\bibitem{AWM03}
C.~Ahn, H.~M. Wiseman, and G.~J. Milburn, ``Quantum error correction for
  continuously detected errors,'' \emph{Phys. Rev. A}, vol.~67, p. 052310,
  2003.

\bibitem{GrW04}
M.~Gregoratti and R.~F. Werner, ``On quantum error-correction by classical
  feedback in discrete time,'' \emph{J. Math. Phys.}, vol.~45, p. 2600, 2004.

\bibitem{SAJM04}
M.~Sarovar, C.~Ahn, K.~Jacobs, and G.~J. Milburn, ``Practical scheme for error
  control using feedback,'' \emph{Phys. Rev. A}, vol.~69, no. 052324, 2004.

\bibitem{HJHS03}
A.~Hopkins, K.~Jacobs, S.~Habib, and K.~Schwab, ``Feedback cooling of a
  nanomechanical resonator,'' \emph{Phys. Rev. B}, vol.~68, no. 235328, 2003.

\bibitem{SROKW02}
W.~P. Smith, J.~E. Reiner, L.~A. Orozco, S.~Kuhr, and H.~M. Wiseman, ``apture
  and release of a conditional state of a cavity qed system by quantum
  feedback,'' \emph{Phys.~Rev.~Lett}, vol.~89, no. 133601, 2002.

\bibitem{DoJ99}
A.~C. Doherty and K.~Jacobs, ``Feedback-control of quantum systems using
  continuous state-estimation,'' \emph{Phys. Rev. A}, vol.~60, pp. 2700--2710,
  1999.

\bibitem{DJJ01}
A.~C. Doherty, K.~Jacobs, and G.~Jungman, ``Information, disturbance and
  hamiltonian quantum feedback control,'' \emph{Phys. Rev. A}, vol.~63, no.
  062306, 2001.

\bibitem{EdB03}
S.~C. Edwards and V.~P. Belavkin, ``On the duality of quantum filtering and
  optimal feedback control in quantum open linear dynamical systems,''
  \emph{Physics and Control, 2003. Proceedings. 2003 International Conference},
  vol.~3, pp. 768--772, 20-22 Aug. 2003.

\bibitem{BEB05}
L.~M. Bouten, S.~C. Edwards, and V.~P. Belavkin, ``Bellman equations for
  optimal feedback control of qubit states,'' \emph{J. Phys. B}, vol.~38, pp.
  151--160, 2005.

\bibitem{Jam04}
M.~R. James, ``Risk sensitive optimal control of quantum systems,'' \emph{Phys.
  Rev. A.}, vol.~69, no. 032108, 2004.

\bibitem{HSM05}
R.~van Handel, J.~K. Stockton, and H.~Mabuchi, ``Feedback control of quantum
  state reduction,'' \emph{IEEE Trans. Automat. Control}, vol.~50, no.~6, pp.
  768--780, 2005.

\bibitem{DHJMT00}
A.~C. Doherty, S.~Habib, K.~Jacobs, H.~Mabuchi, and S.~M. Tan, ``Quantum
  feedback control and classical control theory,'' \emph{Phys. Rev. A},
  vol.~62, no. 012105, 2000.

\bibitem{PK02}
J.~Palao and R.~Kosloff, ``Quantum computing by an optimal control algorithm
  for unitary transformations,'' \emph{Phys.~Rev.~Lett}, vol.~89, no. 188301,
  2002.

\bibitem{TV02}
C.~Tesch and R.~Vivie-Riedle, ``Quantum computation with vibrationally excited
  molecules,'' \emph{Phys.~Rev.~Lett}, vol.~89, no. 157901, 2002.

\bibitem{XYOFR04}
R.~Xu, Y.~Yan, Y.~Ohtsuki, Y.~Fujimura, and H.~Rabitz, ``Optimal control of
  quantum non-markovian dissipation: Reduced liouville-space theory,''
  \emph{J.~Chem.~Phys}, vol. 120, pp. 6600--6608, 2004.

\bibitem{Bel79}
V.~P. Belavkin, ``Optimal measurement and control in quantum dynamical
  systems,'' Institute of Physics, Nicolaus Copernicus University, Torun,''
  preprint 411, 1979.

\bibitem{Bel83}
------, ``On the theory of controlling observable quantum systems,''
  \emph{Automatica and Remote Control}, vol.~44, no.~2, pp. 178--188, 1983.

\bibitem{Bel88}
------, ``Nondemolition stochastic calculus in {Fock} space and nonlinear
  filtering and control in quantum systems,'' in \emph{Proceedings {XXIV}
  Karpacz winter school}, ser. Stochastic methods in mathematics and physics,
  R.~Guelerak and W.~Karwowski, Eds.\hskip 1em plus 0.5em minus 0.4em\relax
  World Scientific, Singapore, 1988, pp. 310--324.

\bibitem{AFL90}
L.~Accardi, A.~Frigerio, and Y.~Lu, ``The weak coupling limit as a quantum
  functional central limit,'' \emph{Commun. Math. Phys.}, vol. 131, pp.
  537--570, 1990.

\bibitem{GaZ00}
C.~Gardiner and P.~Zoller, \emph{Quantum Noise}.\hskip 1em plus 0.5em minus
  0.4em\relax Springer, Berlin, 2000.

\bibitem{Hol82}
A.~S. Holevo, \emph{Probabilistic and Statistical Aspects of Quantum
  Theory}.\hskip 1em plus 0.5em minus 0.4em\relax North Holland, Amsterdam,
  1982.

\bibitem{Par92}
K.~R. Parthasarathy, \emph{An Introduction to Quantum Stochastic
  Calculus}.\hskip 1em plus 0.5em minus 0.4em\relax Birkh\"auser, Basel, 1992.

\bibitem{AGL95}
L.~Accardi, J.~Gough, and Y.~Lu, ``On the stochastic limit of quantum field
  theory,'' \emph{Rep. Math. Phys.}, vol.~36, no. 2-3, pp. 155--187, 1995.

\bibitem{HuP84}
R.~L. Hudson and K.~R. Parthasarathy, ``Quantum {It\^o's} formula and
  stochastic evolutions,'' \emph{Commun. Math. Phys.}, vol.~93, pp. 301--323,
  1984.

\bibitem{GKS76}
V.~Gorini, A.~Kossakowski, and E.~Sudarshan, ``Completely positive dynamical
  semigrouops of n-level systems,'' \emph{J. Math Phys.}, vol.~17, no.~5, pp.
  821--825, 1976.

\bibitem{Lin76}
G.~Lindblad, ``On the generators of quantum dynamical semigroups,''
  \emph{Commun. Math. Phys.}, vol.~48, pp. 119--130, 1976.

\bibitem{Bel80}
V.~P. Belavkin, ``Quantum filtering of {Markov} signals with white quantum
  noise,'' \emph{Radiotechnika i Electronika}, vol.~25, pp. 1445--1453, 1980.

\bibitem{Kal60}
R.~E. Kalman, ``A new approach to linear filtering and prediction problems,''
  \emph{J. Basic Eng}, vol.~82, pp. 34--45, 1960.

\bibitem{KaB61}
R.~E. Kalman and R.~S. Bucy, ``New results in linear filtering and prediction
  theory,'' \emph{J. Basic Eng.}, pp. 95--108, 1961.

\bibitem{Tak71}
M.~Takesaki, ``Conditional expectations in von {Neumann} algebras,'' \emph{J.
  Funct. Anal.}, vol.~9, pp. 306--321, 1971.

\bibitem{Bel92b}
V.~P. Belavkin, ``Quantum stochastic calculus and quantum nonlinear
  filtering,'' \emph{Journal of Multivariate Analysis}, vol.~42, pp. 171--201,
  1992.

\bibitem{Bel92a}
------, ``Quantum continual measurements and a posteriori collapse on {CCR},''
  \emph{Commun. Math. Phys.}, vol. 146, pp. 611--635, 1992.

\bibitem{Bel99}
------, ``Measurement, filtering and control in quantum open dynamical
  systems,'' \emph{Rep on Math Phys}, vol.~43, no.~3, pp. 405--425, 1999.

\bibitem{BeS92}
V.~P. Belavkin and P.~Staszewski, ``Nondemolition observation of a free quantum
  particle,'' \emph{Phys. Rev. A}, vol.~45, no.~3, pp. 1347--1356, 1992.

\end{thebibliography}

\end{document}